\pdfoutput=1

\documentclass[11pt]{article}

\usepackage{ACL2023}

\usepackage{times}
\usepackage{latexsym}
\usepackage{booktabs}

\usepackage[T1]{fontenc}

\usepackage[utf8]{inputenc}

\usepackage{microtype}

\usepackage{inconsolata}

\usepackage{hyperref}
\usepackage{url}
\usepackage{graphicx}
\usepackage{verbatim}
\usepackage{todonotes}
\usepackage{xspace}
\usepackage{multirow}
\usepackage{siunitx}
\usepackage[inline]{enumitem}

\newcommand{\ours}{SPEAR-TTS\xspace}
\newcommand{\fs}{FastSpeech2-LR\xspace}
\newcommand{\Sone}{$\mathcal{S}_1$\xspace}
\newcommand{\Stwo}{$\mathcal{S}_2$\xspace}
\newcommand{\wvbert}{w2v-BERT\xspace}

\title{Speak, Read and Prompt: High-Fidelity Text-to-Speech \\ with Minimal Supervision}

\author{Eugene Kharitonov \\
  \And
  Damien Vincent  \\
  \And Zal\'an Borsos \\
  \And Rapha\"el Marinier \\
  \AND Sertan Girgin \\
  \And Olivier Pietquin \\
  \And Matt Sharifi \\
  \And Marco Tagliasacchi \\
  \And Neil Zeghidour \\
  \AND \\ Google Research \\ \texttt{\{kharitonov,damienv,neilz\}@google.com}
  }

\begin{document}
\maketitle
\begin{abstract}
We introduce \ours, a multi-speaker text-to-speech (TTS) system  that can be trained with minimal supervision. By combining two types of discrete speech representations, we cast TTS as a composition of two sequence-to-sequence tasks: from text to high-level semantic tokens (akin to ``reading'') and from semantic tokens to low-level acoustic tokens (``speaking''). Decoupling these two tasks enables training of the ``speaking'' module using abundant audio-only data, and unlocks the highly efficient combination of pretraining and backtranslation to reduce the need for parallel data when training the ``reading'' component. To control the speaker identity, we adopt {example prompting}, which allows \ours to generalize to unseen speakers using only a short sample of 3 seconds, without any explicit speaker representation or speaker-id labels.
Our experiments demonstrate that \ours achieves a character error rate that is competitive with state-of-the-art methods using only 15 minutes of parallel data, while matching ground-truth speech in terms of naturalness and acoustic quality, as measured in subjective tests.
\end{abstract}

\section{Introduction}
\looseness=-1
Training a text-to-speech (TTS) system typically requires hundreds of hours of parallel data in the form of transcribed utterances. As a consequence, TTS is only available for ``high-resource'' languages. Moreover, the audio generated by such systems is only as diverse as the parallel data that they are trained on, which should contain many speakers, with various accents, of diverse demographics, and heterogeneous recording conditions.
At the same time, for most languages, including low-resource ones, audio-only speech data can be relatively abundant online, present in the forms of audiobooks, podcasts, radio and TV shows.

\looseness=-1In this paper, we investigate how audio-only data can be leveraged to reduce the need for supervision in training TTS systems. We introduce \ours{},\footnote{SPEAR stands for ``\textbf{sp}eak, r\textbf{ea}d and p\textbf{r}ompt''.} a multi-speaker TTS system that can be trained with as little as 15 minutes of parallel data from a single speaker. Moreover, \ours{} can synthesize a new voice using only 3 seconds of speech, without any speaker labels or explicit speaker representation. At its core, \ours{} leverages recent advances in the ``textless'' modeling of spoken language~\citep{Lakhotia2021,Dunbar2021,Polyak2021,Kreuk2021,Kharitonov2022,Nguyen2022,Borsos2022}. These methods represent continuous audio waveforms as sequences of tokens from a finite vocabulary, casting speech generation as a language modeling task. In particular, AudioLM~\citep{Borsos2022} combines two types of discrete tokens: high-level semantic tokens and low-level acoustic tokens, which that can be mapped to audio. 
Using these representations, we cast the TTS problem as a ``translation'' from text transcripts to acoustic tokens with semantic token representations serving as a pivot ``language''~\cite{Utiyama2007}. This way, TTS is reduced to a composition of two sequence-to-sequence (seq2seq) tasks: translating text to semantic tokens, and translating semantic to acoustic tokens.

\looseness=-1
The key benefit of splitting the TTS task into these two sub-tasks is that the supervision needed to learn how to map text into the intermediate semantic token representation (``reading'') and how to produce speech from it (``speaking'') become decoupled. While the ``reading'' stage relies on parallel text-audio data, the audio tokens used to train the ``speaking'' component are produced by self-supervised audio models and therefore can be extracted from a massive amount of unlabeled speech data. 
As a result, the quality and diversity of the generated speech become independent from the available parallel data. 

\looseness=-1
Casting each stage of \ours{} as a seq2seq problem allows us to use standard Transformer models~\citep{Vaswani2017} and makes it easy to tap into the vast pool of ideas developed by the machine translation community to reduce the need for supervision. Specifically, we combine BART/T5-style pretraining~\citep{Lewis2020,Raffel2020} with backtranslation~\citep{Sennrich2016} to significantly reduce the amount of parallel supervision required to train \ours{}.

\looseness=-1
To control the voice used by \ours{} when producing an utterance, we leverage an example prompting mechanism that is closely related to prompting in textual language models~\citep{Brown2020}. Here we condition the ``speaking''  model with an audio clip
representing the target voice, steering it to use this voice when generating the utterance. 
This feature can simplify building controllable multi-speaker TTS systems for languages where only single-speaker parallel data is available.

Modeling speech synthesis with seq2seq models enables using stochastic sampling at inference, which allows generating outputs of diverse quality for the same input. We exploit that to improve the synthesized audio quality by proposing a sampling scheme based on an objective quality metric.

\looseness=-1
Our experimental study on English speech shows that, by combining pretraining and backtranslation over a large dataset --- 551 hours from LibriTTS~\cite{Zen2019} --- with just 15 minutes of parallel data from a single speaker --- LJSpeech~\cite{ljspeech} --- \ours{} (a) generates speech with high fidelity to the input transcript --- CER 1.92\% on LibriSpeech test-clean~\cite{Panayotov2015}); (b) synthesizes speech with diverse voices, (c) reliably reproduces the voice of an unseen speaker, when using a 3 second example from the target speaker;
(d) achieves high acoustic quality, comparable to that of the ground-truth utterances (MOS 4.96 vs.\ 4.92).\footnote{Samples produced
by \ours can be found on the demo site: \url{https://google-research.github.io/seanet/speartts/examples/}.}

\looseness=-1
Overall, our approach to building TTS using massive self-supervised pretraining and backtranslation of discrete speech representations considerably differs from how existing TTS systems are implemented~\cite{Shen2018,Kong2020,Ren2020,Kim2021,Ao2022,valle}, significantly reducing the costs related to data collection and potentially providing high-quality multi-speaker TTS for languages that are not well covered today. 

\section{Discrete Speech Representations} \label{s:representation}
Below we provide a brief overview of the self-supervised audio representations that are essential for \ours.
The combined use of these representations was proposed in AudioLM~\citep{Borsos2022}, which we refer to for a detailed discussion.

\paragraph{Semantic tokens}
The role of semantic tokens is to provide a coarse, high-level conditioning to subsequently produce acoustic tokens. Thus, they should provide a representation of speech where linguistic content --- from phonetics to semantics --- is salient, while paralinguistic information such as speaker identity and acoustic details are removed. To obtain such a representation, we train a self-supervised speech representation model based on \wvbert \citep{Chung2021}. This model combines masked language modeling \citep{bert} and contrastive learning \citep{cpc} to obtain speech representations.
After its training, we run  a $k$-means clustering on the mean-variance normalized outputs of a specific layer. We use the centroid indices as discrete tokens.
\begin{figure*}[t]
\centering
\includegraphics[width=0.8\textwidth,trim={3cm 5cm 2cm 5cm}]{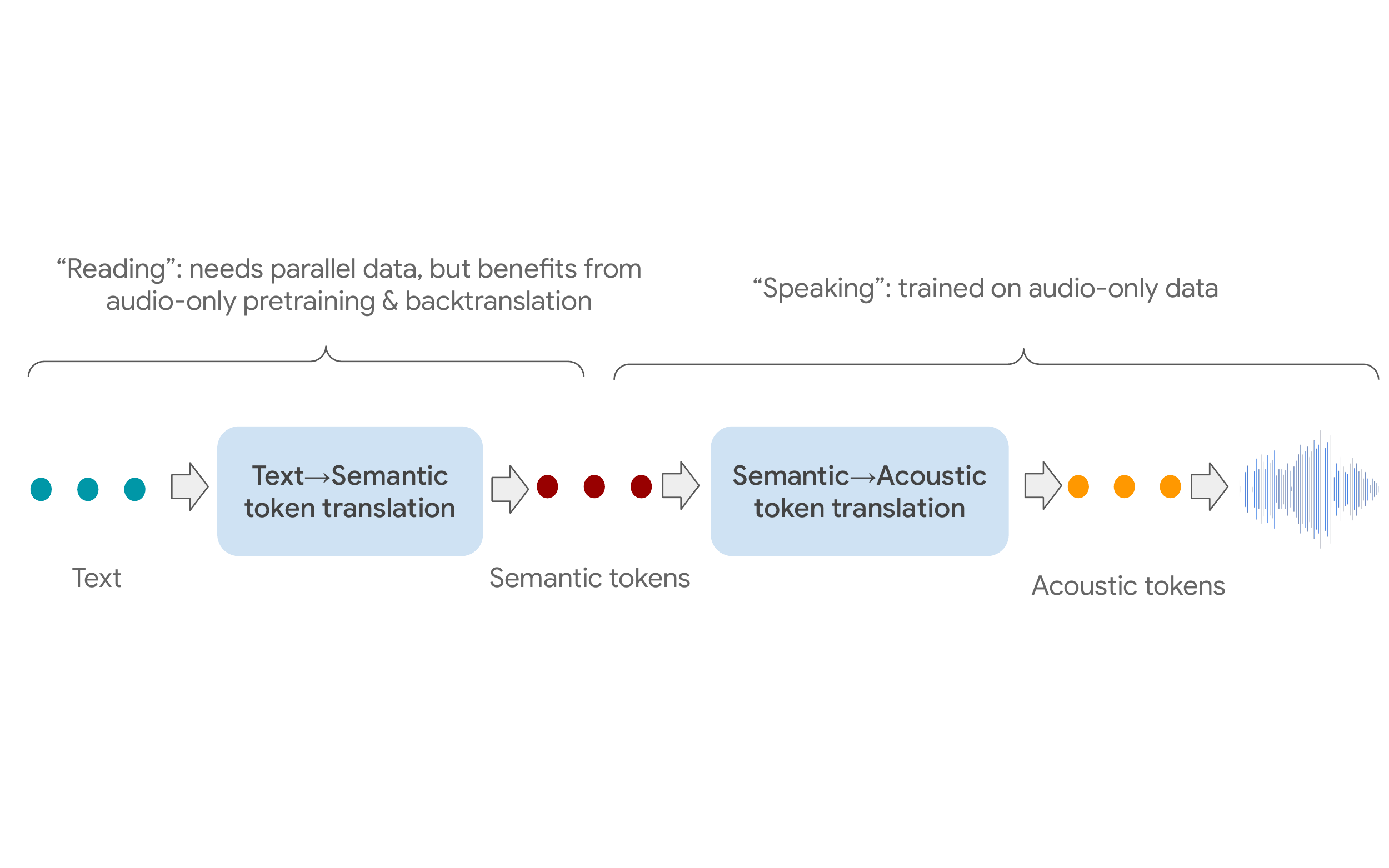}
\caption{\textbf{\ours.} The first stage \Sone{} (``reading'') maps tokenized text to semantic tokens. The second stage \Stwo (``speaking'') maps semantic tokens to acoustic tokens. Acoustic tokens are decoded to audio waveforms.\label{fig:tts_scheme}}
\end{figure*}

\paragraph{Acoustic tokens}
Acoustic tokens are discrete audio representations that provide high-fidelity reconstruction of the acoustic details. We train a SoundStream \citep{Zeghidour2021} neural codec to reconstruct speech while compressing it into few discrete units. SoundStream achieves this goal by adding a residual quantizer to the bottleneck of a convolutional autoencoder. 
To represent the hierarchy of residual quantizers in a sequence, we flatten the tokens corresponding to the different levels by interleaving them~\cite{Borsos2022}.

\section{\ours{} Overview}
\label{s:architecture}
\looseness=-1
\ours{} extends  AudioLM~\citep{Borsos2022} by enabling text as a form of conditioning. \ours{} is organized in two main stages, as illustrated in Figure~\ref{fig:tts_scheme}. In the first stage (\Sone{}), text inputs are translated into a sequence of discrete semantic tokens. The second stage (\Stwo{}) maps semantic tokens into acoustic tokens, which are decoded to speech by the SoundStream decoder~\cite{Zeghidour2021}. This way, \Sone{} learns to map text to the internal representation provided by semantic tokens (``reading''), while \Stwo{} handles the production of speech from this intermediate internal representation (``speaking'').

\looseness=-1
By using semantic tokens as an intermediate representation, we achieve two goals. First, semantic tokens provide a speech representation that encodes mostly phonetic content, with limited prosody and speaker information, bridging the gap between text and acoustic tokens. As a result, our intermediate representation is closer to the text than acoustic tokens are. Thus, it is easier to learn a mapping from text transcripts to semantic tokens than directly between text and acoustic tokens. Second, as both semantic and acoustic tokens are derived from self-supervised models,
the second stage \Stwo{}
can be trained using audio-only data.
This turns out to be extremely beneficial for training \Stwo, as the typical scale of available audio-only data is considerably larger than that of parallel data.\footnote{In the case of English, a large dataset such as LibriTTS has 580h of parallel data~\cite{Zen2019}, while LibriLight contains 60,000h of untranscribed speech~\cite{Kahn2020}.}  
\looseness-1In turn, separating \Sone{} from \Stwo{} allows us to pretrain the former with a denoising pretext task operating on semantic tokens, further harnessing audio-only data.

\begin{figure*}[t]
\centering
\includegraphics[width=0.9\textwidth,trim={2cm 2cm 1cm 0cm}]{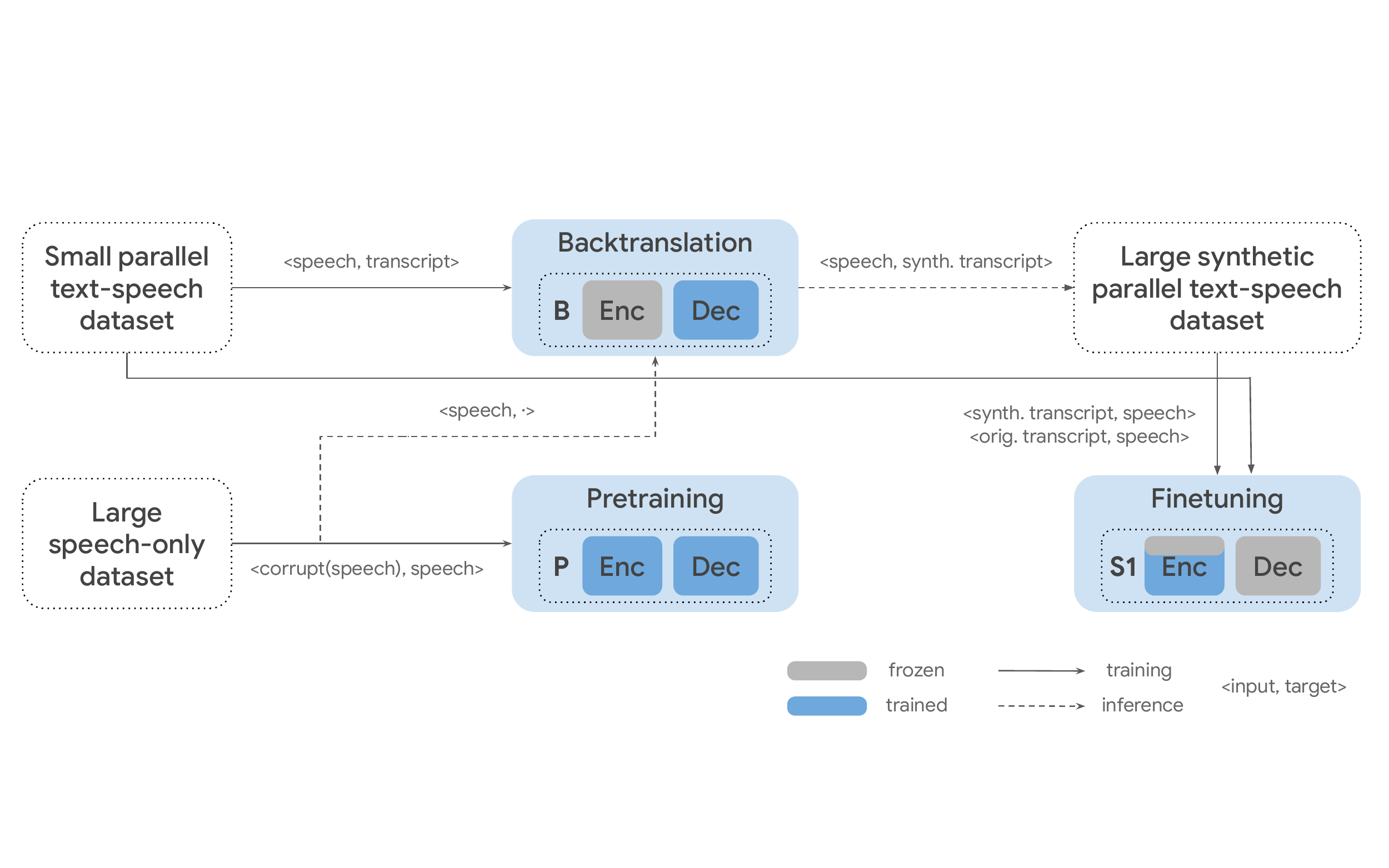}
\caption{\textbf{Training \Sone, combining pretraining and backtranslation}. We start with pretraining an encoder-decoder model $\mathcal{P}$ on a denoising task, using a semantic-token representation of speech-only data. Next, we finetune its decoder to backtranslate (semantic tokens to transcripts) using a small parallel dataset. Then, we use this model to transcribe the speech-only dataset and obtain a synthetic parallel dataset. In turn, this synthetic dataset is used to finetune the encoder of $\mathcal{P}$ for ``translation'' in the forward direction (text transcripts to semantic tokens), along with the original small parallel dataset. \label{fig:stage_one}}
\end{figure*}

Similar to AudioLM~\cite{Borsos2022}, it is possible to add an optional third stage,  with the goal of improving quality of the synthesized speech by predicting acoustic tokens corresponding to fine residual vector quantization levels (Appendix~\ref{s:bandwidth-extension}).

\section{\Sone: Improving Supervision Efficiency}
The first stage \Sone{}  maps tokenized text into semantic tokens. We use parallel text-semantic tokens data to learn this mapping. We start with a text-audio TTS dataset and extract semantic tokens from audio. As a result, \Sone{} is reduced to a seq2seq task, that can be implemented by encoder-decoder or decoder-only Transformer architectures~\citep{Vaswani2017,Raffel2020}.

Training Transformer seq2seq models can require substantial amounts of parallel data, which can be extremely scarce for low-resource languages. In the following, we discuss two approaches used to alleviate this limitation: target domain pretraining (Section~\ref{ss:pretraining}) and backtranslation (Section~\ref{ss:backtranslation}).

\subsection{Pretraining}
\label{ss:pretraining}
We take inspiration from BART and T5 and pretrain an encoder-decoder Transformer on a denoising pretext task~\citep{Lewis2020,Raffel2020}. In this task, the model is provided with a corrupted version of an original semantic token sequence and the goal is to produce the corresponding uncorrupted token sequence.

Typical corruption methods include random substitution, deletion and masking of individual tokens or entire spans of tokens~\citep{Raffel2020}. In preliminary studies, we observed that deleting individual tokens independently with a constant probability works better than other alternatives.

After pretraining the model $\mathcal{P}$ on the denoising task, we finetune it for the \Sone task. To achieve this, we freeze the upper layers of the encoder and all parameters of the decoder, excluding the parameters used in the decoder-encoder cross-attention layers, and update the lower layers of the encoder. The exact number of layers to tune is a hyperparameter.

\begin{figure*}[t]
\centering
\includegraphics[width=0.75\textwidth,trim={1cm 4cm 4cm 2cm}]{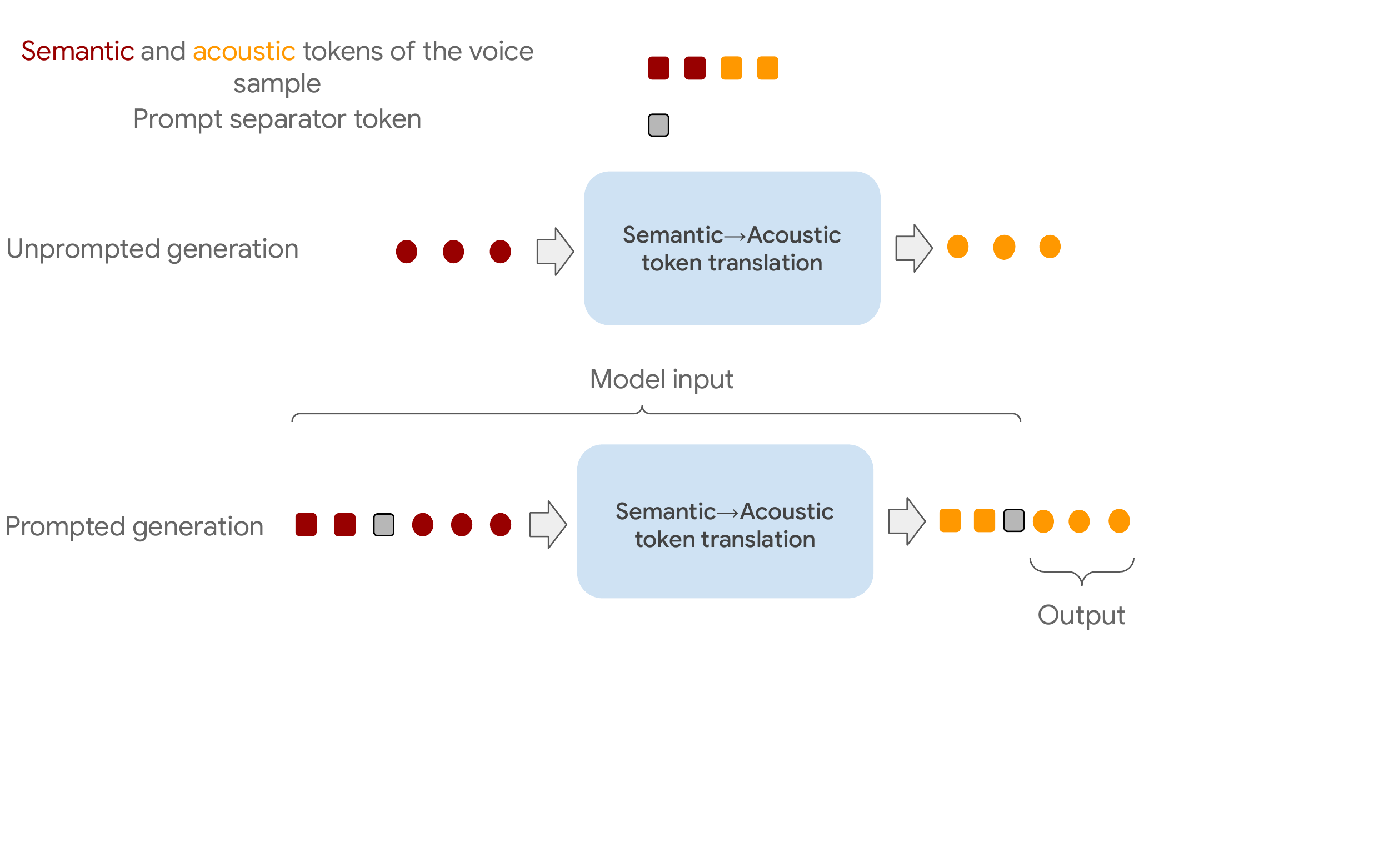}
\caption{\textbf{Controlling generation with example prompting in \Stwo.} 
For prompted generation, we concatenate token sequences in the following order: semantic tokens from the prompt, semantic tokens from the target, acoustic tokens from the prompt. Then, the model generates acoustic tokens corresponding to the semantic tokens from the target, while preserving the voice and speaking conditions in the acoustic tokens from the prompt.
\label{fig:prompting}}
\end{figure*}

\subsection{Backtranslation}
\label{ss:backtranslation}
The same text sequence can be rendered as audio in multiple ways, with varying voice, accent, prosody, emotional content, and recording conditions. This one-to-many  relationship makes the text-to-speech problem highly asymmetric ---  unlike text translation, where, for example, English-French translation is roughly equally hard in either direction.
Thus, it is very attractive to use backtranslation~\cite{Sennrich2016,Edunov2018}, i.e.,  to use the available parallel data to train a speech-to-text model and use it to generate synthetic parallel data from an  audio-only corpus.

The two-stage architecture of \ours{} is particularly suitable for backtranslation as it can be implemented as translation between semantic tokens and text. The benefits are two-fold: (a) a reduction in the computational complexity due to never dealing with raw audio or long acoustic token sequences,\footnote{In our setup, an acoustic token representation of an utterance is at least 6$\times$ longer than its semantic token counterpart.} and (b) the ability to leverage the same semantic token-level pretraining (Section~\ref{ss:pretraining}) when training the ``backward''-direction model, from semantic tokens to text transcripts.

In order to obtain a backtranslation model, we start from the same pretrained model $\mathcal{P}$ as above. However, this time we freeze the encoder and only finetune the decoder. Afterwards, we transcribe audio-only data using this model. Next, we use the synthetically generated parallel data to train the first stage of the TTS system, which, in turn, {is also obtained via finetuning} another copy of $\mathcal{P}$ (see Section~\ref{ss:pretraining}). After finetuning on the synthetic data, we continue finetuning on the original parallel data.\footnote{Another option is to train on a mixture of synthetic and original data as in~\cite{Edunov2018}, which introduces mixture weights as another hyperparameter.} In Figure~\ref{fig:stage_one} we illustrate this combined pretraining and backtranslation process.

\section{\Stwo{}: Controlling the Generation Process}
\label{s:generation}
The second stage model \Stwo{} maps semantic tokens into acoustic tokens.  To train this stage, we extract pairs of sequences of semantic and acoustic tokens from each utterance in an audio-only dataset. 
Next, we train a Transformer model to perform seq2seq translation between the two token sequences.  The second stage generates utterances with randomly varying voice, tempo, and recording conditions, reproducing the distribution of the characteristics observed in the training data. As training of \Sone and \Stwo are decoupled, this diversity of speech generated by \Stwo is preserved also when \Sone is trained on a single-speaker dataset.

\looseness-1To control the characteristics of the speaker's voice, we combine two findings from  AudioLM~\cite{Borsos2022}: (a) whenever the speech prefix is represented solely by semantic tokens, AudioLM generates continuations by sampling a different random voice each time, (b) however, when conditioning also includes acoustic tokens, AudioLM maintains the voice characteristics captured by the acoustic tokens when generating the continuation.
In contrast to AudioLM, we explicitly incorporate this ability during training, as illustrated in Figure~\ref{fig:prompting}.
During training, we randomly select two non-overlapping windows of speech from each training example, from which we compute sequences of semantic and acoustic tokens. We consider one of the windows as the prompt and the other as the target output. Next, we concatenate the sequences in the following order: (a) semantic tokens from the prompt, (b) semantic tokens from the target, (c) acoustic tokens from the prompt, and (d) acoustic tokens from the target. During training of \Stwo, (a)-(c) are used as prefix and the model learns to generate the target acoustic tokens (d), preserving the speaker identity captured by the acoustic tokens from the prompt. At inference time, (a)-(c) are provided as input, and (d) is generated autoregressively. 

Importantly, a special separator token is added at each segment boundary to inform the model about the expected discontinuity. This prevents boundary artifacts, which are sometimes generated when no separator is used. Note that the text transcript of the prompt is not needed.

\looseness-1The speech samples generated by \Stwo might contain some background noise, since this is typically present in the training data. We consider two methods to control the noise level in the synthesized speech at inference time. First, in the case of prompted generation, it is possible to select prompts containing cleaner speech. Second, we can use a stochastic sampling (e.g., temperature sampling), generate multiple sequences for the same input and then use a no-reference audio quality metric to select the samples containing the least amount of noise. To this end, we use a MOS estimator model similar to DNSMOS~\cite{Reddy2021}.

\section{Experimental Setup}

In this section we introduce the datasets, metrics and baselines used in our experimental study.

\subsection{Training and validation data}
\label{ss:data_training}
\paragraph{Acoustic and semantic tokens:} 
We use LibriLight~\citep{Kahn2020}
to train the self-supervised representation models (SoundStream and w2v-BERT) as well as the $k$-means used to discretize w2v-BERT embeddings into semantic tokens. We use the largest unlab-60k split
of LibriLight that contains around 60,000 hours of English audiobooks read by more than 7,000 speakers. 
\paragraph{First stage $\mathcal{S}_1$:}
\looseness-1To experiment in the low-resource regime, we train \Sone{} on LJSpeech~\cite{ljspeech}, a single-speaker dataset containing 24 hours of parallel data. By using LJSpeech as the only source of parallel data, we also show that our method generalizes to multiple speakers, even if the parallel training data itself contains only a single speaker.
Since LJSpeech does not specify a canonical train/dev/test split, we follow~\citet{Liu2022tts,Liu2020} and randomly select 300 utterances as development and another 300 utterances as test set (30 minutes each), using the rest as training data. To simulate scenarios in which very limited data is available, we uniformly sample subsets of 12, 3, 2, 1 hours, 30, and 15 minutes from the training set. As an indicative figure, the 15 minute subset contains around 21k semantic tokens and 2k words.

\paragraph{Pretraining:} To pretrain a model on the sequence corruption task (Section~\ref{ss:pretraining}), we extract semantic tokens from LibriLight~\citep{Kahn2020}, since pre-training only requires audio data.

\paragraph{Backtranslation:}
In our experiments with backtranslation, we use LibriTTS~\cite{Zen2019} as a source of unlabelled speech (ignoring transcripts). We pool all training subsets of LibriTTS
to obtain an audio-only dataset containing 551 hours of speech.
Using LibriTTS as a source for audio-only data for backtranslation allows us to compare \ours{} with \Sone trained on original and backtranslated LibriTTS transcripts.

\paragraph{Second stage \Stwo{}:}
To train \Stwo, we extract pairs of semantic and acoustic token sequences from LibriLight~\citep{Kahn2020}.

\subsection{Evaluation data}
\label{ss:data_evaluation}

We use LibriSpeech test-clean~\citep{Panayotov2015} to measure the character error rate (CER)~(see Section~\ref{ss:metrics}). As LJSpeech only contains sequences shorter than 10 seconds, we filter out sequences longer than that from LibriSpeech test-clean. As a result, we obtain 2,007 utterances, with a total duration of approximately 3 hours. Importantly, LibriSpeech test-clean has no intersection with any training or validation data we used.

\subsection{Preprocessing}\label{ss:preprocessing}
To prepare the data for training, we unroll standard abbreviations used in LJSpeech. Next, we apply the {G2p\_en} phonemizer~\citep{g2p}. After removing the lexical stress information from its output, we obtain a string representation in a vocabulary of 47 tokens (39 phonemes from the CMU Dictionary, whitespace, and punctuation).

Since we cannot expect that a phonemizer is universally available in low-supervision scenarios, in Appendix~\ref{s:characters} we experiment with grapheme inputs.

\subsection{Metrics} \label{ss:metrics}
We are interested in the following desired properties of \ours:
\begin{itemize}
\vspace{-1mm}
    \itemsep-0.1em
    \item Generated speech should adhere to the input;
    \item It should provide voice diversity even when \Sone{} is trained on single-speaker data;
    \item When prompted with an utterance from an unseen target speaker, \ours{} should synthesize speech that matches their voice;
    \item Generated speech should be of high quality.
\end{itemize}
Below we discuss the metrics used to assess whether those properties are satisfied. 
\paragraph{Character Error Rate (CER)} We transcribe the utterances synthesized by \ours using an in-house ASR system and we evaluate the faithfulness to the input transcript by measuring the character error rate (CER).
We use the LibriSpeech test-clean dataset~\citep{Panayotov2015} to calculate CER, since it requires minimal postprocessing to be compared to the output of the adopted ASR system. As a reference, on the original ground-truth audio, CER is equal to 0.98\%.

\paragraph{Voice diversity} To measure the voice diversity within a set of synthesized speech utterances, we apply a speaker classifier that assigns one speaker per utterance and we measure the entropy of the empirical distribution of the detected speakers across all utterances.
We use the same speaker classifier as~\citet{Borsos2022}, which is trained on a union of LibriSpeech train-clean-100 and test-clean containing 251 and 40 speakers, respectively, and computes predictions over a set of 291 speaker classes. We provide more details in Appendix~\ref{s:speaker_classifier}.

\paragraph{Voice preservation} When prompting the model with a short utterance, we evaluate the consistency of the speaker voice between the prompt and the generated speech. To this end, we use the same speaker classifier as above and measure how often the speaker label predicted from the generated speech matches the one predicted from the prompt.

\paragraph{Quality} We rely on human judgments to evaluate the perceived quality of \ours{} by collecting Mean Opinion Scores (MOS). In this context, human raters listen to individual audio segments and rate their audio quality and speech naturalness on a scale from Poor (1) to Excellent (5).

\subsection{Baselines}\label{ss:baselines}
As our main baseline, we consider a system explicitly trained to target the low-supervision scenario. Namely, we use a modification of FastSpeech2~\cite{Ren2020}, which is a non-autoregressive model that uses auxiliary duration, pitch, and energy predictors. Specifically, in our experiments we consider the adaptation to the low-resource setting by~\citet{Pine2022}. The model takes as input the phoneme representation of the text and predicts a spectrogram, which is then vocoded with HiFi-GAN~\cite{Kong2020}.   We denote this modification as \fs. In a subjective evaluation reported by \citet{Pine2022}, \fs trained on 1 (3) hour(s) of parallel data performed on par with an open-source implementation of Tacotron2~\cite{Shen2018} trained with 10 (24) hours of parallel data.
We use checkpoints trained on 15 minutes, 30 minutes, 1 hours, 3 hours, and 24 hours subsamples of LJSpeech that were shared by the authors.\footnote{\url{https://github.com/roedoejet/FastSpeech2_ACL2022_reproducibility}}

We also compare \ours{} to VALL-E~\cite{valle}, a recent TTS system that demonstrates state-of-the-art results in zero-shot voice adaptation. Similarly to \ours, it is capable of voice transfer using a 3 second voice prompt. VALL-E maps the input text to coarse acoustic tokens, and uses a non-autoregressive refinement stage to predict fine-grained acoustic tokens. VALL-E is trained on an ASR-transcribed version of LibriLight~\cite{Kahn2020}, containing roughly 60,000 hours of parallel data. Since the model is not publicly available, the comparison is based on the samples provided on its demo page.

\section{Hyperparameters \& Training details}\label{s:hyperparams}
\subsection{Discrete Speech Representations} \label{ss:discrete-speech-repr}
We follow the setup of AudioLM~\cite{Borsos2022} to compute both semantic and acoustic tokens, with a few differences. The semantic tokens are obtained by quantizing the embeddings returned by the 7th layer of \wvbert{} using a codebook of size 512.
As a result, 1 second of audio is represented by 25 semantic tokens with a vocabulary size of 512, resulting in an equivalent bitrate of $25 \times \log_2{512} = 225$ bit/s. We remove sequentially repeated semantic tokens, as done in~\citet{Lakhotia2021,Borsos2022}.

We extract acoustic tokens from a SoundStream neural codec~\citep{Zeghidour2021} with 3 quantization levels, each with a codebook of size 1024. We use a vocabulary with $3 \times 1024$ unique tokens and represent each frame as a flat sequence of tokens, interleaving the first, second, and third quantization layers, respectively. As a result, 1 second of audio is represented by 50 Hz $\times$ 3 = 150 acoustic tokens, an equivalent bitrate of 1500 bit/s.

\subsection{First stage (\Sone{})}
In all experiments, we use the Adafactor optimizer~\citep{Shazeer2018} with inverse square-root learning rate decay. As a regularization method, we use label smoothing with the smoothing parameter set to 0.1, except in the case of pretraining, when a large amount of data is available.

\paragraph{Pretraining} The pretraining task is configured so that the probability of deleting individual tokens is set to 0.6. This parameter was selected via grid search inspecting the validation accuracy of \Sone after finetuning.
We apply dropout with probability equal to 0.5 and set the batch size to 256. We ran the pretraining for 1M updates and used the resulting checkpoint $\mathcal{P}$ in all our experiments. As the architecture, we use T5-Large~\cite{Raffel2020}, which is a 24 layer encoder-decoder seq2seq model (see Appendix~\ref{s:t5_details}).

\paragraph{Finetuning} The same pretrained checkpoint $\mathcal{P}$ is finetuned for different purposes (Figure~\ref{fig:stage_one}). In all cases we perform a grid search on the dropout rate (\{0.1, 0.3, 0.5\}) and the number of layers to finetune, selecting the combination with the highest validation accuracy (with teacher-forcing). 
More specifically, when finetuning on ground-truth parallel data (as an ablation), we freeze both the upper layers of the encoder and the entire decoder, while updating the weights of the encoder embeddings and the lower layers. The number of the lower layers to tune is searched in \{4, 6, 8\}.
When finetuning on synthetic parallel data, we search over the number of the encoder's lower layers to be finetuned in \{4, 6, 8, 10, 12, 24\}. Next, we finetune the lower 4 layers of the encoder on the original parallel data (to avoid overfitting when very little data is available). Finally, when finetuning the decoder for backtranslation, we finetune $N$ top and $N$ bottom layers, with $N \in \{2, 3, 4, 12\}.$ During finetuning, we select the checkpoint with the best validation accuracy.

\paragraph{Training from scratch}
As an ablation experiment, we train \Sone from scratch, experimenting with different variants of T5 architectures~\cite{Raffel2020}, depending on the amount of data available.
We adopt a decoder-only model without causal masking on the input sequence~\cite{Raffel2020}, which led to better results in our preliminary experiments.
We perform a grid-search on the following hyperparameters: dropout probability \{0.1, 0.3, 0.5\}; architecture size (T5-small or T5-base); the number of layers (T5-small: 2, 4, 6, 8; T5-base: 4, 6, 8, 12). Further details are in Appendix~\ref{s:t5_details}.

\subsection{Second stage (\Stwo{})}
\looseness-1 For \Stwo, we use a 12-layer decoder-only Transformer model, with each layer having 12 heads with dimensionality 64, embedding dimensionality of 768, and FFN size of 2048. The optimizer and the learning rate schedule are the same as for \Sone.

\subsection{Inference}\label{ss:inference}
\looseness-1We use beam search to sample from \Sone and temperature sampling to sample from \Stwo{}. 
This combination ensures faithfulness to the transcript while enabling more diverse and natural sounding speech. We use a beam size equal to 10, as larger values do not lead to improvements in CER but are more computationally expensive. When generating backtranslation data we re-use the settings of $\mathcal{S}_1$, without running any additional hyperparameter search. 
For \Stwo, we experiment with sampling temperatures $T \in \{0.50, 0.55, ..., 0.95, 1.0\}$ and select $T=0.75$ which minimizes the CER on the LJSpeech validation dataset. In this case, the \Sone model is trained on synthetically generated parallel data obtained by backtranslation, with the backtranslation model trained on the 15 minute split of LJSpeech.

To control the noise levels in the synthesized speech, we employ the sampling technique (Section~\ref{s:generation}) where we sample $n_s$ audio utterances corresponding to the input and select the one that has highest quality according to a no-reference audio quality model similar to DNSMOS~\cite{Reddy2021}. We set $n_s$ to 3, as a trade-off between audio quality and computational cost (Appendix~\ref{ss:resampling}).

\section{Experiments}

\label{sec:experiments}
We evaluate \ours{} along several dimensions. First, we measure the faithfulness of the generated speech to the input transcript, for different training scenarios and amounts of parallel data available~(Section~\ref{ss:fidelity}). Then, we observe that \ours{} is able to generate speech that is more diverse in voices than the parallel data used during training (Section~\ref{ss:prosody}). Finally, we show that \ours is able to successfully control the speaker voice, without any degradation in terms of fidelity to the transcript (Section~\ref{ss:prompting}).

\begin{table*}[t]
    \caption{\textbf{Intelligibility} of \ours{} and our baselines, depending on the training scenario and the amount of parallel data available from LJSpeech. We measure CER (\%, lower is better) on LibriSpeech test-clean. $\pm$ indicates 95\% CI obtained by bootstrap.``$\times$'' indicates models that produce unintelligible speech.} \label{tab:ljspeech_cer}
    
  \centering
  \resizebox{0.9 \textwidth}{!}{
    \begin{tabular}{lcccccccc}
    \toprule
 & & \multicolumn{4}{c}{\ours}  \\
 & &
 
\multirow{2}{*}{\begin{tabular}{c}Training \\ from scratch (a)\end{tabular}} & \multirow{2}{*}{Pretraining (b)}  & \multicolumn{2}{c}{Backtranslation} \\
Parallel training data & \fs  & & &  from scratch (c) &  pretraining (d) \\
 \midrule

  24 h & 1.99$_{\pm 0.20}$ & 3.67$_{\pm 0.21}$ & 2.38$_{\pm 0.13}$ & 2.26$_{\pm 0.14}$ & 2.06$_{\pm 0.12}$ \\
  12 h & - &  4.31$_{\pm 0.28}$ & 2.54$_{\pm0.14}$ & 2.27$_{\pm 0.14}$ & 2.03$_{\pm 0.12}$ \\ 
  3 h & 2.52$_{\pm 0.25}$ & 20.1$_{\pm 0.74}$ & 3.07$_{\pm0.15}$ & 2.21$_{\pm 0.12}$ &  2.01$_{\pm 0.12}$ \\ 
  2 h & - &  24.7$_{\pm 0.71}$ & 3.73$_{\pm0.17}$ & 2.22$_{\pm 0.13}$  & 2.09$_{\pm 0.12}$ \\ 
  1 h & 2.74$_{\pm 0.27}$ & $\times$ & 5.51$_{\pm0.21}$ & 2.23$_{\pm 0.13}$  & 2.16$_{\pm 0.13}$ \\ 
  30 min & 3.18$_{\pm 0.28}$ & $\times$ & 21.3$_{\pm0.43}$ & 2.52$_{\pm 0.15}$  & 2.20$_{\pm 0.12}$ \\ 
  15 min & 4.90$_{\pm 0.34}$ &  $\times$ & $\times$ & 2.88$_{\pm 0.19}$  & 2.21$_{\pm 0.12}$\\ 
    \bottomrule
    \end{tabular}
    }

\end{table*}

\subsection{Intelligibility and Supervision Efficiency} \label{ss:fidelity}
When evaluating \ours, we consider the following training settings for \Sone:
\begin {enumerate*} [label=\itshape(\alph*\upshape)]
    \item training from scratch using parallel data;
    \item finetuning the pretrained checkpoint $\mathcal{P}$ using parallel data;
    \item finetuning the pretrained checkpoint $\mathcal{P}$ to obtain the backtranslation model and then training the forward model from scratch on the synthetically generated data;
    \item same as (c), but both the backward and the forward models are obtained by finetuning  $\mathcal{P}$ with an additional finetuning of the forward model on the original parallel data.
\end {enumerate*} 

Table~\ref{tab:ljspeech_cer} reports the main results in terms of CER, as a proxy for the intelligibility of the generated speech. We observe that when decreasing the amount of parallel data, training from scratch (a) results in very high error rates. Conversely, thanks to pretraining (b), \ours{} maintains a relatively low CER ($\le 4\%$), when using as little as 2 hours of parallel data. This is similar to the CER achieved with 24 hours, but without pretraining. Backtranslation (c) has a general positive impact, especially when the amount of parallel data is reduced, achieving a CER of 2.88\% with only 15 minutes. By combining backtranslation with pretraining (d), the CER is further decreased to 2.21\% with the same amount of parallel data. This
indicates that having a fixed decoder is useful to cope with the noisy nature of the synthetically generated training data obtained via backtranslation. As a result, \ours{} trained on 3 hours (with pretraining and backtranslation) achieves the same CER that can be observed when training from scratch on the original transcripts of LibriTTS-train, that is, 551 hours of parallel data (see Appendix~\ref{s:libritts}). 

We also compare \ours{} to \fs, observing that when using 24 hours of parallel data, both systems perform approximately on par (\fs: 1.99\% vs.\ \ours{}: 2.06\%). However, as the amount of parallel data is reduced, CER of \fs increases very rapidly. As a result, there is a significant gap when only 15 minutes are available, that is, \fs: 4.90\% vs.\ \ours{}: 2.21\%. 

In conclusion, the combination of pretraining and backtranslation allows \ours{} to synthesize speech that adheres to the input transcript, even with as little as 15 minutes of parallel data.

%

\subsection{Voice diversity}
\label{ss:prosody}

\ours{} 
is capable of generating utterances using diverse voices, including speakers not seen in the parallel data. For example, when using the LJSpeech dataset~\cite{ljspeech} as the source of parallel data, the model generates multiple different voices, despite the fact that this dataset contains a single speaker. In the following experiments, we quantitatively measure the voice diversity of the generated speech. 

To this end, we train \Sone on parallel datasets characterized by a different number of speakers and verify that the diversity of the synthesized voices remains stable. We consider 1 speaker (LJSpeech), 61, 123 and 247 speakers (from LibriTTS). Namely, we use the full LibriTTS train-clean-100, which contains 247 speakers and two its subsets with approximately 1/2 and 1/4 of the speakers. We use transcripts from LibriSpeech test-clean.

Table~\ref{tab:speaker_entropy} illustrates how the ground-truth speech naturally becomes less diverse in terms of voice variability (from 7.68 to 2.55), as the number of speakers is decreased (from 247 to 1). Note that the LJSpeech voice is out-of-domain for the speaker classifier used, so the measured voice variability is non-zero. Instead, for \ours{}, voice variability is not significantly affected by the number of speakers (min: 6.16, max: 6.28) and significantly higher than \fs (6.11 vs.\ 0.66). 

This experiment demonstrates that the variety of voices synthesized by \ours is independent from the number of distinct speaker voices contained in the parallel data used for training \Sone.

\begin{table}[t]
    \caption{\textbf{Voice diversity (bits).} We measure the entropy of the empirical distribution of the voices detected by a speaker classifier.}
    \label{tab:speaker_entropy}  
  \centering
  \resizebox{0.9 \columnwidth}{!}{
    \begin{tabular}{lccccc}
    \toprule
 & LJSpeech & \multicolumn{3}{c}{LibriTTS}  \\
 \# speakers & 1 & 61 & 123 & 247 \\ 
\midrule
Ground-truth & 2.55 & 5.82 & 6.71 &  7.68 \\
\ours{} & 6.11 & 6.22 & 6.16 & 6.28 \\
\fs & 0.66 & - & - & - \\

  \bottomrule
    \end{tabular}
    }
\end{table}

\subsection{Prompted generation}\label{ss:prompting}

\ours{} is able to control the speaker voice via example prompting, as described in Section~\ref{ss:prompting}. We evaluate \ours{} in a \textit{zero-shot} scenario, in which the voice used for prompting was never seen by \Sone or \Stwo at training  and \Stwo has to reproduce its characteristics from a single prompt example. Specifically, we fix \Sone{}, using the model trained on 15-minutes of LJSpeech and we consider all 40 speakers from LibriSpeech test-clean as target speakers. For each speaker, we randomly select 5 speech prompts with duration of 3 seconds each and transcripts from the same dataset. For each speech prompt and text transcript, we repeat synthesis 5 times and average metrics across the generated samples. 

\looseness-1Table~\ref{tab:libritts_speakercond_baseline} reports the speaker accuracy, that is, how often the same voice is detected in both the prompt and the generated speech. We observe top-1 accuracy equal to $92.4\%$ showing that the prompting allows \ours{} to preserve the speaker voice. Also, the synthesized voice is stable when repeating inference, as captured by a low value of voice variability (0.41 bits). Moreover, we observe that with prompted generation \ours achieves a CER equal to 1.92\%, which is lower than without prompting (2.21\%). We believe that this improvement is due to using cleaner recordings for prompts, which steers the \Stwo model to produce cleaner speech and consequently reduce ASR errors.

We also compare the voice preservation abilities of \ours{} with those of VALL-E~\citep{valle}. Following the methodology of ~\citet{valle} we compute the cosine similarity between embeddings computed from the prompt (encoded and decoded with SoundStream) and from the generated speech, using a publicly available speaker verification system based on WavLM~\citep{wavlm}.\footnote{\url{https://github.com/microsoft/UniSpeech/tree/main/downstreams/speaker\_verification\#pre-trained-models}, ``WavLM large'' model.} This is the same model used by~\citet{valle} which makes our measurements directly comparable with scores reported in their paper. From the results reported in Table~\ref{tab:speaker_sim}, we observe that \ours{} significantly outperforms YourTTS~\citep{Casanova2022} (0.56 vs.\ 0.34) and almost matches the speaker similarity of VALL-E (0.58), despite being trained with 240,000$\times$ less parallel data.

\begin{table}[t]
        \caption{\textbf{Voice preservation in prompted generation (classifier accuracy).} 
        The \Sone model is trained on 15 min of parallel data. 
        } \label{tab:libritts_speakercond_baseline}
  \centering
  \resizebox{0.9 \columnwidth}{!}{
  \begin{tabular}{lcccc}
  \toprule
   CER (\%) & \multicolumn{2}{c}{Speaker accuracy (\%)} & Voice diversity (bits) \\
   & {\small top-1} & {\small top-3} & \\
   \midrule
   1.92   & 92.4 & 98.1 & 0.41 \\
  \bottomrule
  \end{tabular}
  }

\end{table}

\begin{table}[t]
        \caption{\textbf{Comparing voice preservation with baselines (cosine similarity).} Results for YourTTS and VALL-E are taken from~\cite[Table 2]{valle}. 
        } \label{tab:speaker_sim}
  \centering
  \resizebox{0.9 \columnwidth}{!}{
  \begin{tabular}{lcc}
  \toprule
   Model & Parallel training data & Cosine similarity \\
   \midrule
   YourTTS   & $\sim$ 600 h & 0.34  \\
   VALL-E   & 60,000 h & 0.58 \\
   \ours   & 15 min & 0.56 \\
  \bottomrule
  \end{tabular}
  }

\end{table}

\section{Subjective Evaluation}\label{ss:subjective}
\label{ss:mos_low}
Ultimately, we resort to subjective tests with human raters to compare the quality of \ours{} with the baselines and with ground-truth natural speech. We focus on the scenario with minimal supervision and use the \Sone model that is trained with the 15 minute LJSpeech~\cite{ljspeech} subset. As baselines, we use the \fs models~\cite{Ren2020,Pine2022} trained on 15 minutes, 1 hour, and 24 hour subsets of LJSpeech. 

To ensure that the evaluation sentences are not part of the training set of \ours or the \fs models, we extract sentences from an audiobook chapter released in 2022, read by the same voice as in LJSpeech.\footnote{\url{https://librivox.org/predecessors-of-cleopatra-by-leigh-north/}, \S 10.}  This chapter was published later than any of the datasets we use. We extract 20 sentences from it, each with duration between 3 and 11 seconds, for a total of 133 seconds. We take transcripts for those sentences in the text of the corresponding book.\footnote{ \url{https://www.gutenberg.org/cache/epub/58236/pg58236.txt}} We provide the transcripts in Table~\ref{tab:sentences} in Appendix.

The baselines are TTS systems trained to generate a single voice. To ensure a fair comparison, we prompt \Stwo with utterances extracted from the LJSpeech dataset, so that \ours{} generates speech with the same voice.  To this end, we randomly select 3s speech samples from LJSpeech and filter out samples that have more than 1s of silence, using the remaining as prompts.

\looseness-1Samples are presented to raters one-by-one, and raters are asked to judge the audio quality and speech naturalness on a scale from Poor (1) to Excellent (5). Before starting, the raters were provided with example utterances for each grade. Each audio sample is evaluated by 20 raters. For each treatment, we average all scores to compute the Mean Opinion Score (MOS).

Table~\ref{tab:mos_low} reports the results of the subjective tests. We observe that \ours achieves considerably higher quality than the baselines, even when the latter use more parallel data during training. The MOS score achieved by \ours (4.96) is comparable to the one obtained for the ground-truth speech (4.92), confirming the high quality of the generated speech, despite the fact that the model was trained only on 15 minutes of parallel data.

\begin{table*}[t]
    \caption{\textbf{Mean Opinion Score (MOS) evaluation.} All compared systems are trained on subsets of LJSpeech~\cite{ljspeech}. $\pm$ indicates 95\% CI obtained by bootstrap.\label{tab:mos_low}}
  \centering
  \resizebox{0.8 \textwidth}{!}{
    \begin{tabular}{lcccccc}
    \toprule
 &    \multicolumn{3}{c}{\fs} & \ours & Ground-truth \\
Parallel training data  & 15 min & 1 h & 24 h & 15 min & - \\
\midrule

MOS &  $1.72_{\pm 0.04}$ & $2.08_{\pm 0.04}$ & $2.11_{\pm 0.04}$ & \textbf{4.96}$_{\pm 0.02}$ & $4.92_{\pm 0.04}$ \\
  \bottomrule
    \end{tabular}
    }

\end{table*}

We also compare \ours{} and {VALL-E}~\cite{valle} in  a small-scale subjective test using the examples provided on its demo page.\footnote{\url{https://valle-demo.github.io/}, ``More Samples''.} 
These examples are generated by combining 8 transcripts with 3 prompts each, resulting in 24 speech utterances. Using the same instance of \ours described above (with \Sone trained with 15 minutes of single-speaker LJSpeech), we synthesize 24 utterances using the same transcripts and prompts and conduct a subjective test with the same protocol described above. Table~\ref{tab:valle} shows that,  on these examples, \ours achieves considerably better naturalness and higher speech quality (MOS 4.75) than VALL-E (3.35), despite using considerably less supervision (15 min of parallel data \& 1 speaker vs.\ approximately 60,000 hours of parallel data spoken by over 7,000 speakers).

\begin{table}[t]
    \caption{\textbf{Mean Opinion Score (MOS) evaluation for prompted generation.} Prompts for both systems and samples for VALL-E are taken from the demo page of VALL-E. $\pm$ indicates 95\% CI obtained by bootstrap.} \label{tab:valle}
  \centering
  \resizebox{0.8 \columnwidth}{!}{
    \begin{tabular}{lcccccc}
    \toprule
System &  VALL-E & \ours (15 min) \\
\midrule
MOS  &  $3.35_{\pm 0.12}$ & \textbf{4.75}$_{\pm 0.06}$ \\
  \bottomrule
    \end{tabular}
    }

\end{table}

\section{Related Work}
\subsection{Discretized Speech Processing}
The work of \citet{Lakhotia2021} on generative spoken language modeling (GSLM) pioneered the use of language models on discretized speech representations. The main tasks \citet{Lakhotia2021} focuses on are unconstrained speech generation and speech continuation. Their work became a foundation for a range of applications and extensions, including emotion transfer \cite{Kreuk2021}, prosody \cite{Kharitonov2022} and dialog \cite{Nguyen2022} modeling.  \ours{} is related to AudioLM~\citep{Borsos2022}, a recent development in this line of work 
that achieves a superior quality in spoken language modeling as well as a high audio quality.  
\subsection{Low- and semi-supervised TTS}

\looseness-1Being able to leverage audio-only data is one of the distinct features of \ours. Guided-TTS, proposed by \citet{Kim2021}, is another TTS system that is capable of doing this. At its core, Guided-TTS combines (a) a denoising diffusion probablistic model (DDPM) that learns to model audio-only data, and (b) a phoneme classifier that guides the generative process towards producing an utterance with a desired transcript. Guided-TTS~2~\citep{Kim2022} extends Guided-TTS by allowing speaker adaptability either via finetuning or in a zero-shot manner, using a 10 second speech sample processed by a dedicated speaker embedding module. Another adaptable DDPM-based TTS system was proposed by \citet{Levkovitch2022}, which uses the classifier guidance mechanism to steer generation towards a particular voice in a zero-shot manner.

\looseness-1In contrast to \ours{}, the above works rely on a stronger supervision: (a) Guided-TTS uses a phoneme classifier that is trained on LibriSpeech 960, (b) Guided-TTS~2 relies on a pre-trained speaker verification system. Conversely, \ours uses an intuitive and parameter-less prompting mechanism which does not require any speaker labels.

\looseness-1\citet{Liu2020} combine a sequential autoencoder with vector quantization and temporal segmentation mechanisms to learn a phoneme-like discrete speech representation, along with a seq2seq model that maps these representations to phonemes. Similarly to \ours{}, this system can be trained with almost no supervision, however the generated speech is single-speaker only and of much lower quality than ground-truth audio ($2.33$ vs $4.81$ in their experiments). This is unlike \ours{} which despite minimal, single-speaker supervision can generate speech from arbitrary voices while matching the quality of ground-truth speech.

Next, there is a body of research that exploits availability of unpaired texts.
Backtranslating audio-only data, as done by \ours, can be thought of using an ASR system to generate training data for TTS. \citet{Tjandra2017} proposed to
train both ASR and TTS
simultaneously, with TTS reconstructing the waveform based on the ASR output and ASR recognizing audio, synthesized by TTS.
\citet{Chung2019} discussed a set of approaches for pretraining the Tacotron TTS system, that includes per-frame autoregressive pretraining of the decoder and pretraining word embeddings for the encoder. \citet{Ao2022} proposed SpeechT5, a system that can combines text- and audio-only data for pretraining. 

\subsection{Prompted Audio Generation}
When a sentence is prepended by an emotional prompt, expressed in a plain English, e.g.~[\textit{I am really sad, }] Tortoise TTS~\cite{Betker2022} synthesizes text in a sad-sounding voice.

AudioLM~\cite{Borsos2022} demonstrates a voice-prompting ability where an acoustic token prefix forces the model to maintain the speaker characteristics and recording conditions in the prompt, while generating a speech continuation. We extend the prompting capabilities of AudioLM by proposing prompt-aware training of \Stwo.

\citet{valle} propose VALL-E, a TTS system that allows prompt-based conditioning of the synthesized voice and emotion. In contrast to the two-stage architecture of \ours, VALL-E predicts an equivalent of acoustic tokens directly from a phoneme representation of a text. As a result, the transcript of the prompt is required, which can be challenging e.g.\ if the prompt is noisy. This is unlike \ours{} which only prompts the model with self-supervised audio tokens, and thus does not require the corresponding transcript. Another difference is the amount of the parallel training data used: VALL-E is trained on the 60,000 hours of ASR-transcribed LibriLight~\cite{Kahn2020}. Sections \ref{ss:prompting} and \ref{ss:subjective} show that \ours{} provides similar zero-shot prompting abilities with much higher audio quality, even when trained with only 15 minutes of parallel data.

\section{Conclusions \& Future work}
In this work, we introduce \ours{}, a multi-speaker TTS system that has two features setting it apart. First, it only requires a minimal amount of parallel data to be trained, i.e.\ it can synthesize speech with high fidelity and voice diversity when trained on as little as 15 minutes of parallel data coming from a single speaker. Second, \ours{} is able to synthesize speech maintaining voice characteristics of a previously unseen speaker using a 3-second long voice example. 

\looseness-1These capabilities originate from harnessing abundant audio-only data. The key component that unlocks the usage of such data is the hierarchical discrete representation of speech that combines high-level semantic tokens with low-level acoustic tokens. Using these representations, \ours{} casts the TTS problem as a composition of two sequence-to-sequence tasks, ``reading'' (from tokenized text to semantic tokens) and ``speaking'' (from semantic tokens to acoustic tokens). 

\ours{} uses audio-only data in three ways: (a) to train the ``speaking'' model, such that the hard task of speech generation benefits from large-scale data, (b) as a domain for pretraining a model that is further used as a foundation for text-to-semantic tokens and semantic tokens-to-text models, and (c) to generate synthetic parallel data for backtranslation.

\looseness-1Our experimental study on English data (Section~\ref{sec:experiments}) shows that by combining audio-only data from LibriTTS~\cite{Zen2019} with 15 minutes of parallel data sampled from LJSpeech~\cite{ljspeech}, \ours{} achieves intelligibility comparable to that of an adapted \fs trained on the entire 24 hours of LJSpeech (CER 1.92\% vs.\ 1.99\% on LibriSpeech test-clean~\cite{Panayotov2015}). Simultaneously, even when trained on parallel data from a single speaker, \ours{} synthesizes speech with diverse voices (Section~\ref{ss:prosody}).

Next, our experiments in Section~\ref{ss:prompting} show that \ours{} can maintain voice characteristics of a previously unseen speaker, in a zero-shot manner, with high accuracy. Indeed, our measurements indicate that by taking a 3 second-long voice example for a speaker from LibriSpeech test-clean, \ours{} achieves 92.4\% accuracy on maintaining the voice when synthesizing held-out text transcripts, according to our speaker classifier. Moreover, when measuring speaker similarity between prompts and generated speech, \ours obtains a cosine similarity of 0.56, which is close to the score reported for VALL-E~\cite{valle} and significantly higher than the score of YourTTS~\cite{Casanova2022} (0.58 and 0.34, respectively).

\looseness-1 Subjective evaluations of speech naturalness show that \ours has significantly higher quality than a strong single-voice baseline even when trained with 96$\times$ less parallel data (MOS 4.96 vs.\ 2.11). Moreover, the MOS score of \ours is on par with the natural speech (4.92). When comparing quality of the speech synthesized in a zero-shot voice transfer task, \ours obtains a MOS that is considerably higher than VALL-E  (4.75 vs.\ 3.35), with 240,000$\times$ less data.

\looseness-1We  believe  our  work  on  high-quality  TTS  with limited supervision (quantity- and quality-wise) paves the way for enabling TTS technology for communities that are currently excluded due to speaking ``low-resource'' languages and dialects. Another exciting potential application that can be unlocked by \ours{} is allowing people with speech impairments to use old recordings of their own voice to communicate orally. At the same time, we admit that our initial study has certain limitations and could be misused (Sections~\ref{s:limitations} \& \ref{s:impact}).

We believe that applying our findings to building a TTS system for truly low-resource languages and further reducing the need for supervision are the main directions for further work.

\section{Limitations}\label{s:limitations}
While our motivation is to enable high-quality, diverse, and controllable TTS for low-resource languages, we started our investigations with English, which allowed us to address the problem using a collection of well-studied datasets.

\looseness-1Next, we rely on LibriLight~\cite{Kahn2020} as our audio-only dataset which provides a sufficiently diverse set of audio. However, LibriLight only contains audio samples at 16 kHz, hence \ours{} requires an additional step to synthesize speech at a higher sampling rate (Appendix~\ref{s:bandwidth-extension}). In addition, LibriLight contains audio of a lower quality than curated datasets on average. However, these are not limitations of \ours{}, but rather are limitations of the data we used. Moreover, the quality of \ours{} could be improved by changing the neural codec used to produce acoustic tokens, with no change to \Sone{} and \Stwo{}.

Finally, the flexibility of \ours{} comes from relying on relatively large Transformer models that require substantial computing resources for training and inference. We believe this can be addressed separately by model distillation and quantization~\cite{Polino2018,Fan2020}.

\section{Broader Impact}\label{s:impact}
We believe our work on high-quality TTS that requires very limited supervision (quantity- and quality-wise) can be an important stepping stone for enabling this core speech technology for communities that are currently underserved by TTS solutions due to speaking ``low-resource'' languages, i.e., languages that do not have large parallel corpora required for training deep learning models.
Even for high-resource languages, such as English, the ability to harness untranscribed data for speech generation can enable producing speech in accents and dialects that are currently uncovered by the existing TTS systems.
Another exciting potential application provided by \ours{} is allowing people with speech impairments to use recordings of their own voice to prompt \ours{}.

\looseness-1At the same time, we acknowledge that the ability to mimic a voice can have numerous malicious applications, including bypassing biometric identification and for the purpose of impersonation~\cite{Delgado2021,Casanova2022}. Thus it is crucial to put in place safeguards against the misuse and, as an initial step, we verify that speech produced by \ours{} can be reliably detected by a classifier with an accuracy of 82.5\% on a balanced dataset (see Appendix~\ref{s:spoof}). In the future, one can explore other approaches for detecting synthesized speech, e.g.\ audio watermarking.

\section*{Acknowledgements}
The authors are grateful to Ron Weiss and Matthieu Geist for their feedback on a draft of this paper. We also thank Aidan Pine for helping us to obtain and run checkpoints from \citet{Pine2022}.

\bibliography{custom}
\bibliographystyle{acl_natbib}

\newpage
\appendix

\section{Bandwidth extension: from 16 to 24 kHz}\label{s:bandwidth-extension}
\looseness=-1
While relying on LibriLight as our unpaired dataset allows for modeling a diverse set of speakers and conditions in \Stwo, this dataset contains only 16 kHz audio, whereas 24 kHz audio is preferable in many TTS applications. We provide a simple approach via bandwidth extension that enables \ours to generate speech at 24 kHz, while still being able to benefit from the diversity of LibriLight.

We cast bandwidth extension as a sequence-to-sequence task of mapping tokens produced by the SoundStream codec at 16 kHz (Section \ref{ss:discrete-speech-repr}) to the tokens produces by a SoundStream codec at 24 kHz. We train the latter on LibriTTS \citep{Zen2019} with 4 residual vector quantizer layers, a codebook size of 1024 per layer and 50 Hz embedding rate, resulting in a 2000 bit/s codec. To create the training data for this task, we extract SoundStream token sequence pairs from LibriTTS: the target tokens are produced by the 24 kHz codec on the target audio sample, and the input tokens are produced by the 16 kHz codec on the audio sample, after applying a lowpass filter with random cutoff frequencies between 5 and 8 kHz.

Since the sequence-to-sequence formulation of bandwidth extension fits easily into our framework, we train a T5-small encoder-decoder on the task. We note that the training data for this stage is two orders of magnitudes smaller than for \Stwo (LibriTTS vs LibriLight), so with this approach we can benefit at the same time from the acoustic diversity of a large, but low resolution dataset and the quality of a small, but high resolution dataset.

\section{Controlling audio quality by sampling} \label{ss:resampling}

\begin{table}[t]
  \centering
  \resizebox{ \columnwidth}{!}{
    \begin{tabular}{cccccc}
    \toprule
\# samples, $n_s$ & 1 & 2 & 3 & 5 & 10  \\
 \midrule
CER & $2.10_{\pm 0.07}$ & $1.99_{\pm  0.07}$ & $1.93_{\pm  0.07}$ & $1.90_{\pm 0.06}$  & $1.90_{\pm 0.07}$ \\

Audio Quality & $3.68_{\pm 0.44}$ & $3.86_{\pm 0.31}$  & $3.94_{\pm 0.26}$ & $4.02_{\pm 0.21}$ & $4.11_{\pm 0.18}$ \\
  \bottomrule
    \end{tabular}
    }
        \caption{\textbf{Controlling audio quality by resampling and its effect on CER.} We measure the fidelity and quality of utterances produced by \ours{} depending on the number of sampling steps. CER is calculated on LibriSpeech dev-clean, audio quality is measured on MOS scale by a modification of the DNSMOS model~\cite{Reddy2021}. $\pm$ indicates one  Standard Error of the Mean (SEM).} \label{tab:resampling}

\end{table}

\looseness=-1
As discussed in Section~\ref{s:generation}, without prompting, the quality of audio produced by \ours{} matches that of the training data used to train \Stwo{}. As the LibriLight dataset~\citep{Kahn2020} contains audiobooks read by volunteers using their personal equipment, so the quality of the recordings varies a lot.

Here we verify that the sampling technique proposed in Section~\ref{s:generation} allows us to control the quality of the generated speech and study how it affects the recognition error by the used ASR system. In this experiment, for each phoneme input in LibriSpeech dev-clean, we sample $n_s$ times from \ours{} ($n_s \in \{1,2,5,10\}$) and select the sample that has the highest MOS estimate, returned by a modification of the DNSMOS model~\cite{Reddy2021}. We use the selected example for calculating CER.

We report the results of this experiment in Table~\ref{tab:resampling}. We observe that increasing $n_s$ leads to a higher estimated quality. Moreover, higher audio quality allows \ours{} to achieve lower CER. Based on the results in Table~\ref{tab:resampling}, we use $n_s = 3$ in all our experiments, as a trade-off between the computational complexity and the estimated quality estimate.

\section{Training \Sone on LibriTTS}\label{s:libritts}
\looseness-1In this Section, we study intelligibility of \ours with \Sone trained on LibriTTS. We generally use the same hyperparameter grids as in experiments with LJSpeech~\cite{ljspeech} that are reported in Section~\ref{s:hyperparams}. However,  as  LibriTTS is larger than LJSpeech, we also experiment with encoder-decoder models. For the largest training subset of LibriTTS (551h), we also experimented with T5-Large-sized encoder-decoder and  decoder-only architectures (24 layers). For encoder-decoder models, we always set the numbers of layers in the encoder and the decoder to be equal. 

\looseness=-1Table~\ref{tab:libritts_cer} reports CER for two variants of \ours{}: with \Sone trained from scratch and starting from the pretrained checkpoint $\mathcal{P}$, the same as used in the main experiments. We consider three subsets of LibriTTS~\cite{Zen2019} of different sizes (54, 241, and 551 hours).
First, we notice with the largest subset (551h), \ours reaches a low error rate of 2.04\% and, in this case, pretraining provides virtually no improvement. However, with less paired data, pretraining is increasingly important:  it starts to play a role when 241 hours of paired data available and becomes strongly beneficial when training on 54 hours of paired data (CER 2.61\% vs.\ 2.13\%).

\begin{table}[t]
  \centering
  \resizebox{0.9 \columnwidth}{!}{
    \begin{tabular}{ccccc}
    \toprule
 Dataset size & Training from scratch & Pretraining \\
 \midrule

  551 h & 2.04 & 2.01 \\
  241 h & 2.08 & 1.92	 \\
  54 h & 2.61 & 2.13 \\
  \bottomrule
    \end{tabular}
    }
        \caption{\textbf{CER of \ours{} on LibriSpeech test-clean, when training on LibriTTS.} We measure the fidelity of \ours{} depending on the training regime.} \label{tab:libritts_cer}

\end{table}

\section{Speaker classifier}\label{s:speaker_classifier}
We use the same speaker classifier as \citet{Borsos2022}, which is a convolutional network that takes log-mel spectrograms as its input. The spectrograms are calculated with a window size of 25ms, A hop length of 10ms and have 64 mel bins. The network contains 6 blocks, each cascading convolutions with kernels of 3x1 and 1x3. Each block is followed by a ReLU non-linearity and batch normalization~\citep{batchnorm}. The per-block numbers of channels are [64, 128, 256, 256, 512, 512]. The classifier has an input span of 1 second and, to classify a longer utterance, we run a sliding window with a hop length of 250 ms and average predictions across the windows.

\section{Detecting synthesized speech}\label{s:spoof}
In this section we demonstrate that speech generated by \ours{} can be successfully distinguished from real human speech. To this end, we use the classifier trained to detect speech generated by AudioLM~\cite{Borsos2022}. It uses the same architecture as the speaker classifier (Appendix~\ref{s:speaker_classifier}) and was trained to discriminate LibriSpeech train-clean-100~\cite{Panayotov2015} examples, compressed with SoundStream~\cite{Zeghidour2021}, against AudioLM generated speech.

To assess how effective this classifier is on the speech that \ours{} synthesizes, we iterate over examples in LibriSpeech dev-clean and, from each example, generate two utterances: (a) one by synthesising text using \ours, and (b) one by re-synthesising the ground-truth audio via acoustic tokens.\footnote{We do not compare against uncompressed ground-truth audio as this task is trivial for the classifier by allowing it to focus on superficial coding artifacts, thus making it easier to bypass.}
We observe that on this set of samples, our classifier attains an accuracy of 82.5\% on discriminating generated vs.\ natural speech. We believe that this result can be further improved by training the classifier directly on the output of \ours{}.

\section{Architecture details}
\label{s:t5_details}
We report parameters for the Transformer layers we used in Table~\ref{tab:arch_sizes}.

\begin{table}[t]
\centering
  \resizebox{ \columnwidth}{!}{
\begin{tabular}{lccccc}
\toprule
 & Embed.\ dim. & FFN dim.  & Head dim. & \# heads \\
 \midrule
T5-small & 256 & 512 & 64 & 6 \\
T5-base & 768 & 2048 & 64 & 12 \\
T5-large & 1024 & 2816 & 64 & 16 \\

\bottomrule 
\end{tabular} 
}
\caption{\textbf{Architecture details.} We report details for T5-small, T5-base, and T5-large layers. The number of layers used is defined by a grid search (see Section~\ref{s:hyperparams}). \label{tab:arch_sizes}}
\end{table}

\begin{table}[t]
  \centering
  \resizebox{1 \columnwidth}{!}{
    \begin{tabular}{lccccccc}
    \toprule
 & \multicolumn{7}{c}{Parallel training data} \\
 & 24 h & 12 h & 3 h & 2 h & 1 h & 30 min & 15 min \\
  \midrule
Phonemes &  2.06 & 2.03 & 2.01 & 2.09 & 2.16 & 2.20 & 2.21 \\
Graphemes &  1.79 & 1.79 & 2.13 & 2.27 & 2.46 & 2.71 & 3.45 \\
    \bottomrule
    \end{tabular}
    }
    \caption{\textbf{CER (\%) of \ours{} using a grapheme-based text representation.} \ours is trained with pretraining and backtranslation, using 15 minute subset of LJSpeech as parallel data. \label{tab:ljspeech_cer_chars}}
\end{table}

\section{Evaluating grapheme-based \ours{}}
\label{s:characters}
In the main text, we report evaluation results for a variant of \ours{} trained on phoneme representations of text. In some cases, in particular for low-resource languages, a phonemizer might not be available. Hence, we complement our experimental study by evaluating \ours{} trained on grapheme-based representation of transcripts. We report results in Table~\ref{tab:ljspeech_cer_chars}. On comparing these results with Table~\ref{tab:ljspeech_cer}, we observe that having phoneme-based representation brings strong benefits when very little parallel data is available (e.g., 3.45\% vs.\ 2.21\% with 15 minutes). In contrast, with more than 2 hours of parallel data, the benefits of using a phonemizer shrink, and with 12 hours or more, grapheme-based training outperforms the phoneme-based model.

\begin{table}[t]
  \centering
  \resizebox{0.8 \columnwidth}{!}{
  \begin{tabular}{lcccc}
  \toprule
  Downsample factor  & 1 & 2 & 5 & 10 \\
  \midrule
  CER (\%) &  1.99 & 1.99 & 2.36 & 2.92 \\
  \bottomrule
  \end{tabular}
  }
        \caption{\textbf{CER of \ours on LibriSpeech dev-clean vs.\ \Stwo training data size.} We measure how downsampling LibriLight~\cite{Kahn2020} before training \Stwo affects the CER (\%).} \label{tab:stwo_data}
\end{table}

\section{Influence of the data size on \Stwo}
In this experiment, we measure how sensitive \Stwo is to the amount of data used to train it. To this end, we downsample LibriLight~\cite{Kahn2020} by factors of 1, 2, 5, and 10 before training \Stwo models. All models share the same architecture and are trained for the same number of updates and we select the checkpoint with the highest validation accuracy. Next, we combine the selected checkpoints with \Sone trained on LibriTTS~\cite{Zen2019} (with pretraining) and measure intelligibility of \ours on LibriSpeech dev-clean. We report results in Table~\ref{tab:stwo_data}. We notice that reducing the data size 5x starts to affect the performance.

\begin{table*}[t]
    \centering
    \begin{tabular}{l}
    \toprule
\tiny The computation of his reign probably dates from the time he was first associated with his sister or stepmother in the regal power. \\
\tiny He, like his predecessor, was interested in architecture, builded and added to the temples and showed individual taste in his additions.  \\
\tiny Many, many years were occupied in its preparation. \\
\tiny She evidently did not inherit her mother’s characteristics and possibly did not live any great length of time. \\
\tiny There is also a kneeling statue of him, in later life, holding a globular vase in his hand. \\
\tiny We knew less of her than of almost any of the queens, that she continued the royal line and her name seems but brief record of her. \\
\tiny This stands between the two extended paws, on one of which the king’s name has been found inscribed. \\
\tiny Dreams seem to have borne a special art in the family history. \\
\tiny Hence the young king was considered in a sense the son of the god. \\
\tiny Indeed, it was to this last that he owed his wife, for it was on a hunting expedition that he encountered and fell in love with her. \\
\tiny To paint men brown and women yellow was the rule, but to this there were occasional exceptions.
\tiny Perhaps a middle ground may come nearest to the truth. \\
\tiny It is rarely that the name of an Egyptian sculptor is preserved, but this case is an exception.
\tiny The stone is of a yellowish brown color and very difficult to work. \\
\tiny Says one visitor, the surface of the statues was originally beautifully polished. \\
\tiny These sublime sketches in stone are an artist’s work. \\
\tiny The sounds are said by some authorities to have been heard during a period of two hundred and twenty years. \\
\tiny She lived many years after her husband, whose reign was brief, lasting not more than eight or nine years. \\
\tiny He is described as amiable and generous, and showed deference and strong affection both for mother and wife. \\
\tiny He seems among the most pleasing of the Egyptian kings.     \\
\bottomrule
\end{tabular}
    \caption{Transcripts used for the subjective evaluation.}
    \label{tab:sentences}
\end{table*}

\end{document}